# Thoughts on heavy-ion physics in the high luminosity era: the soft sector

*Federico Antinori, Francesco Becattini, Peter Braun-Munzinger, Tatsuya Chujo, Hideki Hamagaki, John Harris, Ulrich Heinz, Boris Hippolyte, Tetsufumi Hirano, Barbara Jacak, Dmitri Kharzeev, Constantin Loizides, Silvia Masciocchi, Alexander Milov, Andreas Morsch, Berndt Müller, Jamie Nagle, Jean-Yves Ollitrault, Guy Paic, Krishna Rajagopal, Gunther Roland, Jürgen Schukraft, Yves Schutz, Raimond Snellings, Johanna Stachel, Derek Teaney, Julia Velkovska, Sergei Voloshin, Urs Achim Wiedemann, Zhangbu Xu, William Zajc.*

**Abstract**

*This document summarizes thoughts on opportunities in the soft–QCD sector from high-energy nuclear collisions at high luminosities.*

**Introduction**

For an accelerator-based program with nuclear beams, the main avenues for making further progress include increasing the integrated luminosity, scanning the center of mass energy, varying the beam species, as well as making suitable upgrades to detectors and software. How to combine these tools of experimental exploration depends on the physics challenges. At the LHC and at RHIC, the recent discoveries related most directly to the collective properties of hot and dense matter are arguably in the soft physics sector, and so are some of the main outstanding challenges of the field. In particular, the comparison of recent data on nucleus-nucleus, proton-nucleus d+Au, $^3$He+A, and proton-proton collisions raise fundamental questions about the origin of apparent collectivity in smaller systems, the nature of QCD equilibration processes and – in the long term – the question of how experimental access may be gained to the physical degrees of freedom mediating collectivity.

In general, collective phenomena in the soft physics sector can be sensitive to system size, system shape and energy density. This suggests that their exploration will profit not only from increased statistics, but also from a systematic variation of beam energy and from a comparative analysis of different collision systems. An assessment of our current understanding and of the experimental accessibility of the open fundamental questions in the soft physics sector is thus pressing, since it may impact the prioritization of the nuclear beam programs in the coming years.

A group of 30 physicists met to explore these general questions in the Koyasan monastic center (Japan) during the first weekend of October 2015 following the Quark Matter conference in Kobe. The working title of the meeting was "Soft physics in the high-luminosity era", and its main aim was to assess the current state of the art in understanding the soft physics sector in ultra-relativistic nucleus-nucleus and proton-nucleus collisions, and to ask which physical insights into QCD plasma properties are within experimental reach in the coming years.

The participants met as individual scientists interested in the study of extreme forms of QCD matter via high-energy nuclear collisions. They met without any mandate from a collaboration, or from their scientific community, or from any funding agencies. They also met without any preconceived idea that a document must be written. A previous meeting of this kind (arXiv:1409.2981 [hep-ph]) had led to a number of suggestions and ideas of how the discoveries made until now with A-A, p/d-A and pp collisions at the LHC, RHIC and SPS, could be followed up with more detailed experimentation and focused theoretical exploration. The main purpose of the meeting in Koyasan was to deepen this assessment in the particular sector of soft probes, with a view toward a future high-luminosity era. The following summarizes the consensus view that developed during the meeting.

**Assessing the flow paradigm in AA collisions**

In the soft physics sector, that is for sufficiently low transverse momentum probes, most of the data collected in heavy-ion collisions at RHIC and at the LHC have been interpreted in terms of collective behavior, i.e., the emergence as a function of time of strongly interacting matter developing in response to density and pressure gradients. This interpretive framework has implemented in several independently-developed simulation packages. The models underlying the simulations are based on the assumption that the colliding system reaches approximate local equilibrium early in the collision, on time scales below ½ fm/$c$, and they evolve the subsequent many-body dynamics in terms of relativistic viscous hydrodynamics. The mechanisms responsible for decoupling and hadronizing particle degrees of freedom from this fluid dynamic system follow local hadronization prescriptions with hadronic rescattering codes employed in recent years to extend these by including final state interactions and to constrain the resulting modeling uncertainties.

Viscous relativistic fluid-dynamic simulations that implement the above-mentioned physics are generally successful in describing simultaneously all the most pertinent telltale signatures of collective behavior in nucleus-nucleus collisions. In particular, they account qualitatively and at least semi-quantitatively for:

- The mass-splitting of the slopes of the hadron $m_T$-spectra and of their differential anisotropic flow coefficients $v_n(p_T)$, both of which are indicative of a common flow field carrying all hadron species;
- Non-zero values for multi-particle anisotropic flow cumulants that remove correlations involving only a small number of particles and thus provide evidence that all particles participate in the observed flow;
- Event-plane correlations whose signs, magnitudes and centrality dependence are consistent with the effects of non-linear hydrodynamic evolution between the fluctuating initial state and the finally observed momenta;
- Fluctuation-induced flow factorization breaking resulting from event-by-event fluctuations in the final-state flow pattern can be qualitatively explained for transverse momenta below 2.5 GeV/*c* in terms of a hydrodynamic response to initial density fluctuations;
- Ridge-like long-range rapidity correlations in the anisotropic flow phenomena, which establish that the geometric features and fluctuations causing them and the system's collective response to these fluctuations are both phenomena that, by causality, point back to the very beginning of the collision;
- The recently observed event-plane de-correlations along the rapidity direction, which again can be understood in terms of a collective response to initial-state density distributions whose eccentricity vectors fluctuate along the beam direction from event to event.

The main physics hypothesis supported by these detailed comparisons of relativistic viscous fluid dynamics to data from nucleus-nucleus collisions is the so-called perfect fluid paradigm. It states that the hot QCD matter produced in nucleus-nucleus collisions approaches local equilibrium efficiently on time scales smaller than 1 fm/*c* and that it maintains local equilibrium throughout an explosive expansion stage. These experimentally testable properties raise fundamental questions about the nature of the matter produced in heavy ion collisions and about the accuracy and possible limitations with which the evolution of this matter can be described fluid dynamically. Questions open for further investigations include in particular:

– *The quest for increased precision*: Notwithstanding their success, purely hydrodynamic models have so far not been able to describe all the abovementioned experimental observations simultaneously within their (mostly quite small) experimental errors. Many of the remaining quantitative deficiencies may be overcome by model refinements within the fluid dynamic paradigm. For instance, adding more sophisticated microscopic pre-equilibrium and post-hadronization evolution models to the viscous hydrodynamic core is known to improve the theoretical description. However, it remains to be established whether all residual discrepancies with experiments can be resolved in this way. In particular, the almost equal magnitudes of elliptic and triangular flow measured in very central (impact parameter b ≈ 0) Pb+Pb collisions at the LHC remain unexplained so far, and the theoretical understanding of flow factorization breaking effects remains to be improved.
One current limitation for the precision and predictive power of dynamical evolution models is that they depend upon model parameters related to initial conditions, transport properties of the evolving medium and their temperature dependence, and the interfaces between its different components that cannot (yet) be computed reliably from first principles. Present modeling efforts choose best-motivated values for these parameters, constrained by data that are believed to be most sensitive to them or by theoretical arguments. However, this calls for a more systematic approach to this multi-dimensional parameter optimization problem. On the one hand, some improvements may result from improving theoretical calculations. For instance, the temperature dependence of shear and bulk viscosities, treated currently as a model parameter, is at least in principle calculable from QCD. On the other hand, progress in the foreseeable future will also be facilitated by more systematic multi-parameter scans. A sustained effort along these lines, which requires appropriate resources in manpower and computing, should be a near-term priority.

In summary, there is a clear perspective for testing with increasing precision and against more comprehensive data sets the strongly supported working hypothesis that the matter produced in heavy-ion collisions is a near-perfect liquid. The main deliverable of such a program of fluid dynamic precision is to obtain the most precise model extraction of QCD transport properties that are calculable from first principles in field theory. At the same time, one may hope that this program can be pushed to a level of experimental and theoretical accuracy at which deviations from the perfect liquid paradigm can be identified, and the physics underlying such deviations becomes experimentally accessible.

– *The quest for additional tests of the perfect fluid paradigm:* By now, essentially all sufficiently abundant processes of soft particle production have been characterized experimentally and interpreted in terms of collective phenomena. It is likely, however, that measurements with increased statistics can yield qualitatively novel insights into the range of validity of the perfect fluid paradigm. For instance, heavy flavor hadrons are known to display non-vanishing elliptic flow, but the accuracy of existing measurements is insufficient for quantitative model comparisons and a complete measurement of all flow harmonics and their centrality dependence is missing. Within the perfect fluid paradigm, heavy flavor is thought to be a probe dragged by the fluid rather than being a component of the fluid. This makes the measurement of collective phenomena in heavy flavor hadrons a novel, complementary test bed for the further exploration of the perfect fluid paradigm, and for testing the interaction between the perfect fluid and the hard probes embedded in it. To elucidate the latter question, the ongoing experimental program should establish up to the differential level of event-by-event correlations how heavy flavor flow is correlated to the flow of light flavor hadrons. Given the large existing uncertainties on heavy flavor flow measurements, it is obvious that such correlation studies will require significantly increased statistics. This is one example of how novel important questions about perfect fluidity will become experimentally accessible in the coming years.

- *Strong versus weak coupling description and the fundamental constituents of hot QCD matter:* Over the last decade, mainly due to the availability of novel strong-coupling techniques for the description of non-abelian plasmas with gravity duals, theory has firmly established that strongly-coupled systems reveal naturally the minimal dissipative properties and small equilibration time scales supported by data comparisons to viscous fluid dynamic models. Remarkably, such strongly-coupled systems do not lend themselves to a description in terms of quasi-particles. At variance to these observations, recent advances in the kinetic modeling of hot QCD reveal that similar values for time scales and transport properties may arise also from a sufficiently complete (resummed) perturbative treatment. In this latter framework, the QCD plasma is composed of microscopic degrees of freedom with (quasi)-particle properties. Beyond a mere debate on whether strong or weak coupling techniques provide a more suitable starting point for calculating properties of hot QCD matter, experimental identification of the microscopic constituents of the plasma clearly requires progress on resolving this (apparent) dichotomy between strong and weak coupling descriptions of hot QCD matter.
  While the studies discussed above demonstrate that very strong coupling between plasma constituents is a sufficient condition for perfect fluidity, it remains unclear whether it is a necessary condition. The perfect fluid paradigm follows logically from the strongly-coupled quark-gluon plasma (sQGP) paradigm, but the sQGP paradigm does not follow necessarily from perfect fluidity. The answer to the question of whether the observed fluid dynamic phenomena necessitate a microscopic understanding of plasma properties in terms of a strong-coupling paradigm is pending. It is one of the main deliverables of further developments of QCD transport theory to establish to what extent close to perfect fluid behavior can arise in non-abelian quantum field theories outside a non-perturbative strong-coupling treatment and what would be its experimental hallmarks.

- *Constraining initial conditions and their hydrodynamization (i.e. the approach to a sufficient degree of local equilibrium to be able to use a* *macroscopic description based on hydrodynamics[1]):* At present, the most significant source of uncertainties in the viscous fluid dynamic modeling of heavy-ion collisions is the scarceness of constraints on initialization parameters. Progress on this point is expected from several directions.

Experimentally, the recent runs at RHIC with asymmetric nuclear beams like Cu+Au, U+U and $^3$He+Au have opened novel opportunities for studying collective phenomena by varying in a qualitatively different way the spatial eccentricities of the initial nuclear overlap region. In the coming years, more complete fluid dynamical modeling studies should provide a complete assessment of the extent to which such asymmetric collision systems can furnish complementary relevant constraints on initial data.[2] More generally, the opportunity arising from a future electron-ion collider for constraining input data not accessible in nuclear or hadronic collisions needs to be updated regularly in the light of ongoing fluid dynamic simulations of ultra-relativistic heavy-ion collisions. It is generally agreed that an electron-ion collider is the ultimate machine for constraining nuclear parton distribution functions, and that it provides the cleanest environment for exploring the qualitatively novel regime of non-linear QCD evolution at small momentum fraction *x*, commonly referred to as the color glass condensate (CGC). Beyond these established conclusions, the remaining task is to sharpen in the coming years our view of how fundamental insights in the qualitatively novel region of saturated partonic systems can improve

---

[1] Traditionally, it was assumed that the application of hydrodynamics requires that the system is approximately *thermalized*, i.e., the distribution in the local fluid rest frame is approximately isotropic and thermal, a state that is reached after some thermalization time that most theoretical approaches predict to be of the same order or longer than the lifetime of the medium created in heavy-ion collisions. However, recent advances have relaxed this stringent requirement by demonstrating that there is a significantly shorter "hydrodynamization time" when *anisotropic* relativistic *viscous* hydrodynamics becomes a correct description of the system.

[2] We note as an aside that the ion sources currently available at the LHC do not permit replication of the RHIC collision program with non-identical nuclear beams at TeV energies. However, a complete assessment of the physics opportunities offered by asymmetric collision systems could yield useful input for future discussions about whether the LHC heavy-ion program would profit from a second ion source. An additional ion source at CERN's LEIR has been discussed recently in the context of a program of medical applications.

our final understanding of the dense and rapidly evolving systems created in nuclear collisions.

On the theory side, effective transport theory for partonic degrees of freedom is the preferred framework for evolving a weakly-coupled pre-equilibrium stage towards a fluid dynamic regime. Transport theory is consistent with fluid dynamics at sufficiently long wavelength and sufficiently long time scales but extends beyond the latter by its ability to account accurately for the microscopic physics of a many-body system also in situations where the fluid dynamic gradient expansion breaks down. Recent theoretical developments have demonstrated the ability of QCD transport theory to approach the viscous fluid dynamic behavior on sub-fermi time scales. Transport theory has thus the potential to replace by a theoretically sound dynamical concept the currently used ad-hoc hydrodynamization time scale at which fluid dynamics is initialized. In the coming years, it remains to be established to what extent this important qualitative idea can be made quantitatively controlled and thus phenomenologically applicable. The development of a detailed dynamical framework of how fluid dynamical behavior arises from a far out-of-equilibrium system at weak coupling will also sharpen our understanding of which characteristics of the initial conditions are relevant for the fluid dynamical evolution of the system.

We note that in the strong-coupling limit, holographic calculations have led in recent years to a theoretically consistent picture of how hydrodynamization occurs. However, due to its characteristic scale dependence, the QCD plasma and its equilibration properties may differ from those of the plasmas of non-abelian theories with gravity duals. Therefore, the approach described here aims at understanding directly hydrodynamization in QCD.

**Flow-like phenomena in small systems**

In recent years, surprisingly small but dense systems probed in high multiplicity p+A and pp collisions were found to display flow-like phenomena. In particular, high multiplicity pp and p+Pb collisions at LHC and d+Au, $^3$He+Au and p+Au collisions at RHIC[3] have been shown to feature similar ridge-like structures, $v_2$ anisotropy and, in some of the systems including high-multiplicity pp collisions, even $v_3$ anisotropy as seen

---

[3] In the following we will simply refer to pA and p + Au to represent p(d or $^3$He)A

in collisions between large nuclei. Measuring higher-order cumulants has reinforced the collective nature of the azimuthal anisotropy seen in p+Pb collisions. New pp data indicate that flow-like correlations may well extend all the way down to minimum bias collisions, though differences in the analysis techniques warrant further scrutiny.

Theoretical attempts to describe the observed flow-like phenomena in pA and pp collisions within the same dynamical model framework as used for heavy-ion collisions have been surprisingly successful. This engenders fundamental questions about whether perfect liquid sQGP is also formed in these much smaller systems. Are the flow-like structures observed in the smallest systems only similar in appearance to what one observes in heavy-ion collisions, or are they similar in their physical origin and thus a crucial element for understanding the emergence of collective phenomena? To make progress on these fundamental questions, several confounding factors and several conceivable interpretations need to be scrutinized in the coming years:

- *Uncertainties of fluid dynamic evolution increase with decreasing system size:* On a qualitative level, the flow-like phenomena observed in pp and p+A collisions feature the characteristics predicted by the hydrodynamic picture. The observed mass-dependence of the differential $v_2(p_T)$, mass-splitting of the slopes of the hadron $m_T$-spectra, and non-vanishing higher $v_2$ cumulants have been reproduced in fluid dynamic simulations of small systems. However, predictions for collisions involving proton projectiles and/or targets have exhibited a strong dependence on how the nucleon thickness function and its event-by-event fluctuations are modeled, seriously degrading the predictive power of fluid dynamic models for these systems. While sub-nucleonic details of the fluctuating spatial structure of a nucleon were found to be rather unimportant in heavy-ion collisions, they matter greatly for a quantitative description of the dynamical evolution of pp and p+A collisions. This sensitivity may be an opportunity to study sub-nucleonic degrees of freedom in the nucleon, but is certainly also a challenge that could limit testing the fluid dynamic paradigm in small systems with an accuracy that is comparable to what has been achieved in the study of nucleus-nucleus collisions.

- *Is there a minimal size for the onset of collective behavior? Where*

*does the fluid dynamic picture apply and where and how does it break down?* With decreasing system size, one may expect a transition from collective behavior to free streaming on scales where the mean free path (mean scattering time) of medium constituents becomes comparable to the system size (system lifetime). Experimentally, however, no indications for the existence of such an onset of collective behavior with system size have been identified so far. Signatures of collectivity such as higher-cumulant flow harmonics display a remarkably weak dependence on system size. Understanding the apparent absence of a minimal scale for collective behavior is a central problem of the field for which qualitatively different solutions remain to be explored in more detail in the coming years:

i) In the limiting case of the strongly-coupled paradigm, the physics of the non-abelian medium does not change further with increasing the coupling constant. In this limit, that is studied theoretically e.g. by holographic calculations, the medium does not carry quasi-particle excitations and hence there is no minimal mean free path. As a consequence of not displaying an internal scale except the scale set by local energy density, such a perfect fluid can exhibit fluid-dynamic behavior in arbitrarily small systems if the local energy density is sufficiently large. Along this line of thinking, the absence of a minimal size for the onset of collective behavior is thus not a problem, but a signature of a strongly-coupled system. To test whether such a radical interpretation of existing data is unavoidable, one would like to understand its implications for a broader class of measurements. For instance, the mere fact that jets, albeit quenched, exit the medium may indicate that the notion of a finite mean free path is of physical relevance at least for sufficiently hard processes. (For more details, see the considerations below on the interaction between the perfect fluid and hard probes.)
ii) Along an orthogonal line of thinking, it has been proposed recently that a short-lived, free-streaming partonic system may display phenomena resembling collectivity if supplemented with a hadronization prescription. We caution that to date there is no satisfactorily complete dynamical framework implementing such a picture. However, the mere fact that the current lack of understanding the signs of collectivity in small systems leads to such a discussion indicates the need for a sharper understanding of which features of collective motion require interactions, i.e., a physical mechanism for establishing a collective flow profile and maintaining local equilibrium during the evolution.
iii) The QCD coupling constant is large, but it is finite and scale-dependent. One thus expects deviations from the perfect fluidity of a strongly-coupled plasma to set in "at some scale". It would be a theoretical breakthrough to establish that scale and to predict its experimental consequences. Recent advances in QCD transport theory make it conceivable that one may clarify in the coming years whether the expected deviations from perfect fluidity are quantitatively consistent with the observed flow-like phenomena in the smallest hadronic collision systems.

In general, fluid dynamics is expected to break down when the gradients become so large that the associated characteristic macroscopic length scales of the system become shorter than the microscopic length scales defined by the mean free paths of the microscopic degrees of freedom (or by the local energy density). This is most likely the case during the earliest collision stage, when the system is still too far from local thermal equilibrium, towards the end of the collision when the system becomes dilute and the mean free paths grow to very large values, and possibly near the quark-hadron phase transition, due to long-range correlations arising from critical fluctuations. Recent developments suggest, however, that large first-order terms in this gradient expansion can be absorbed in a modified hydrodynamic framework called anisotropic hydrodynamics (which is based on an expansion around a leading-order local distribution function that is anisotropic in momentum space). The recent discovery of exact solutions of the Boltzmann equation for systems with highly symmetric flow profiles that nevertheless incorporate key features of expanding systems open new possibilities for identifying criteria that must be satisfied for a macroscopic hydrodynamic approach to accurately reproduce the underlying microscopic dynamics. The current state of the art thus offers several avenues for refining the current fluid dynamical modeling of heavy-ion collisions with recent insights from QCD transport theory.

– *Does collective behavior imply signatures in hard probes?* Collective behavior certainly implies re-interactions between degrees of freedom in the collision region. Such interactions are at the origin of the observed medium-modifications of hard processes. It is therefore relevant to understand better in model studies what the formation of a QGP in small systems would imply about measurable jet quenching effects. On the one hand, jet-quenching effects are known to depend not only on the opacity of the medium, but also on the in-medium path length over which this opacity is accumulated. This makes it conceivable that jet-quenching effects in small systems are relatively small. On the other hand, for a mesoscopic system that exhibits near-perfect liquid behavior, the constituents are expected to find scattering partners with probability close to one. Given that the QCD coupling strength changes only logarithmically with momentum transfer, one may thus expect that also hard processes receive measurable modifications from interactions in the plasma. These considerations should motivate more detailed model studies of jet quenching observables as a function of system size. These studies may also help to elucidate the intriguing measurements indicating that the ridge extends above transverse momenta of 10 GeV/$c$ in p+Pb and 6 GeV/$c$ in d+Au.

– *Do the flow-like features in pp originate from separate physics, or are they a crucial element for understanding the emergence of collective phenomena?* We noted already that with decreasing system size, fluid dynamic modeling becomes more sensitive to initial conditions and thus less predictive. Despite these limitations, it is imperative to establish the physical origin of apparent signs of collectivity in small systems. If the origin is almost perfect fluidity, then this will put tight constraints on properties of medium such as the applicability or non-applicability of the notion of mean free path, and the nature of medium excitations. On the other hand, if near-perfect fluidity is not at the origin of the features observed in pp, then it becomes imperative to establish the elementary partonic origin of these features and to establish to what extent these could also be at play in larger collisions systems. In this context, it is in particular important to better establish whether and to what extent small-$x$ saturation effects formulated in the CGC can account for the azimuthally sensitive, long-range correlations observed in pp and p+A. Even if the effects that can be formulated in this theoretical framework should extrapolate with the system size differently from the flow-like features observed in p+A and A+A, refining successful applications of CGC-motivated ideas to correlation measurements in pp and p+A can inform the ongoing discussion of whether flow-like features in pp could originate from separate physics.

- *Do the systematic variations across beam energies, collision centralities, system size and transverse and longitudinal momentum support a fluid dynamical interpretation? With what accuracy can agreement or disagreement with a fluid dynamic interpretation be quantified?* Enabled by recent advances in more sophisticated model building, this question can be addressed in the near future by comprehensive numerical studies of the scaling behavior of viscous hydrodynamic flow phenomena. Since flow is the hydrodynamic response to the fluctuating initial conditions, which at this point cannot be reliably computed from first principles, a comprehensive model-to-data comparison must include an efficient parameterization of the fluctuating initial state, e.g. in terms of a few numbers determining the means and variances of the dominant radial and angular moments of the initial density profile. These parameters and those describing the medium properties can then be optimized through state-of-the-art multi-dimensional parameter fits using Bayesian techniques. While such a numerical approach will not resolve fundamental limitations in constraining the initial conditions, it has the potential to make explicit the sensitivity of different measurements to model uncertainties, and it is thus a key element for a refined interplay between theory and future experiments. This makes the development of a Bayesian approach to data comparison with fluid dynamic models a near-term priority of the field.

- *Are there interpretational frameworks that do not rely on an approach to approximate local equilibrium (as implicitly assumed in fluid dynamic simulations) but still account for the totality of flow-like correlation measurements in small and large systems?* At present, this question is raised in particular by the phenomenological successes of AMPT, a code that aims at formulating weakly-coupled parton dynamics in terms of a small number of scatterings among well-defined quasi-particle quarks. This code successfully predicted several of the experimentally observed flow phenomena, including phenomena in small and dilute systems. Several workshop participants exchanged views about the origin of this apparent success of a non-fluid dynamic approach. But until now none of these arguments have been documented in the literature. For physics conclusions drawn from phenomenologically successful codes to be reliable, the key dynamical elements and the non-dynamical model prescriptions that lead to flow-like features must be identified and scrutinized carefully. We therefore limit the further comments in this write-up to models that have established a reliability thus far. In particular for theoretical consistency, any model of weakly-coupled parton dynamics should ultimately be demonstrated to be consistent with QCD transport theory (that is, at least in principle, calculable from the QCD Lagrangian).

Similar questions about alternative interpretational frameworks have also been raised by recent studies that appear to suggest that it may be possible to delay hydrodynamization until the point of hadronization without degrading the agreement with experimental data. It has also been suggested that collective-flow-like space-momentum correlations may be generated through the color reconnection mechanism between rapidly dispersing colored partons. In general, all such claims need to be scrutinized through detailed and comprehensive studies, and they must be assessed for their internal theoretical consistency. At present, this process is only beginning.

**Observables**

As detailed in the discussion above, recent data from LHC and RHIC have led to a significant sharpening of fundamental questions about collective behavior in nuclear collisions. Motivated by the primary task of further verifying (or falsifying) the perfect fluid paradigm, the last few years have seen a substantial development towards building up adequately sophisticated and comprehensive dynamical models and in identifying conceivable competing interpretational frameworks. In the coming years, these advances need to be paralleled by an empirical approach that aims at constraining our dynamical understanding of apparent collectivity in dense QCD with the most comprehensive and most precise experimental data accessible.

At present, most of the high-precision data indicative of collective behavior are from two-particle correlation measurements. Future comparisons with theory will rely on having data with better statistics collected differentially as a function of transverse momentum, particle type, rapidity, and as a function of system size and beam energy. Multi-particle correlation measurements, used for example in the analysis of collective anisotropic flow and HBT, have so far confirmed and extended our basic understanding of the properties of the created matter. However, they would in principle for a much better characterization of the collision dynamics and particle production if they were accessible more differentially. In addition, a broad class of novel tests of fluid-dynamic behavior arises from cross-correlating various particle correlation data on an event-by-event basis (such as cross-correlating flow coefficients or reaction plane orientations of different harmonic order as a function of centrality, particle species, rapidity, etc.).

The high luminosity era at the LHC provides an opportunity to perform these differential measurements with unprecedented precision. In addition, RHIC's flexibility to collide nuclei of different size and shape at different energies provides a unique means to test important ingredients of the current paradigm. As detailed above, a major goal in the coming years is to gain a better understanding of the weak dependence of apparent signatures of collectivity on system size *and* on the center-of-mass energy. This will most likely require an experimental program that utilizes the existing options to vary beam species and collision energy.

Particle identification, including the characterization of rare particle species, is another essential element in a comprehensive experimental program to explore the perfect-fluid paradigm. The various particles produced during the evolution of the system interact in different ways with the surrounding matter. Photons and electrons, for example, do not interact strongly and provide direct information about the system at the stage when they are produced. Particles containing strange, charm or bottom quarks are expected to have different interaction strengths with the medium. While the production mechanism of some of these particles is not soft, they still provide important constraints on the soft sector since they allow one to study the degree to which different probes interact with the medium and thus participate in the collective evolution. Correlation measurements between these particles themselves or between different types of particles provide a unique opportunity to understand the various stages of the collision differentially.

The above-mentioned different types of measurements and their systematic variations with particle species and beam energy are important for improved comparisons with fluid dynamical models and competing interpretational frameworks. They arise in a well-defined framework of hypothesis-driven research, they accompany advances in fundamental theory and in developing simulation tools, and they are essential for making progress on the fundamental questions listed in the previous sections. One may note, however, that some of the recent experimental findings that motivate this program were not anticipated by theoretical considerations, and this could happen again, engendering additional questions and/or diagnostic tools. In addition to a clear potential for addressing known fundamental questions, it is conceivable that also in the coming years, refined experimental advances in the soft physics sector will uncover surprising lines of thought. While it is impossible to fully assess future developments a priori, we can point to some profound physics questions that are currently discussed in the scientific community, whose relation to the observed features of collectivity is still not clear but whose future experimental study may yield valuable additional insights into the subject:

- *Hadronization of the plasma vs. hadronization of jets:* Hadronization is the dynamical process in which colored partonic degrees of freedom assemble in color-singlet hadronic bound states. In elementary interactions and in hard processes such as jet production, the phase-space density of colored partons is low, limiting severely the possibilities of forming color-singlet states through local processes. Therefore, while the dynamics of hadronization is not fully understood, existing hadronization models are rather well constrained in these situations. However, in the dense partonic systems produced in heavy-ion collisions and in ultra-relativistic pp and p(d)+A collisions, it is conceivable that color-singlets are formed from partons emerging from different partonic production processes as those considered in coalescence models. Besides the well-studied production of strange baryonic states, it is unclear at present if such novel plasma hadronization effects exist and how they could be identified experimentally. There is, however, the expectation that more detailed analyses of the interaction of hard probes with the medium, including the characterization of kinematic and flavor dependencies with increased statistics and as a function of hard triggers, could shed light on this important question that has remained elusive so far.
- *Time scales and effects of hadronic rescattering:* There is increasing evidence that hadronic rescattering following hadronization (of any source) significantly affects final hadronic distributions, and that it therefore must be taken into account for all analyses involving soft hadrons. This "afterburning" is more relevant at higher multiplicity, and it is therefore important to understand it in very high-energy collisions.

A phenomenological framework aimed at constraining spatio-temporal aspects of this latest stage of heavy-ion collisions has been in existence for many years. Hanbury Brown and Twiss interferometry is employed for evaluating the corresponding correlated volume and the lifetime of the system produced in AA collisions. Non-identical hadron correlations contain information about the temporal order of particle emission, and measurements of photons and low-mass dielectrons that occur throughout the dynamical evolution can play the role of complementary chronometers. A broad range of measurements of hadronic resonances and their correlations can thus constrain further developments of hadronic cascade models. On the one hand, this has the potential of reducing model dependencies in fluid dynamical simulation tools. On the other hand, such data also contains information about the dynamical properties of hadronic resonance gas that are interesting in their own right.

- *Polarization:* If local thermodynamical equilibrium is achieved (one of the key assumptions of the hydrodynamic picture), also the spin degrees of freedom should be in local equilibrium. This leads to the prediction of a (weak) polarization of particles along the orbital angular momentum vector in peripheral collisions. The interesting feature of this polarization is that its magnitude is sensitive to the hydrodynamic initial conditions. The $\Lambda$ baryon polarization can be measured from the angular distribution of its decay products, and the comparison between the signs of $\Lambda$ and anti-$\Lambda$ polarization will allow us to discriminate between vorticity and magnetic fields as their origin. To detect small polarization values high statistics is needed, and therefore such measurements will certainly benefit from the high luminosity run.

- *The physics of strong magnetic fields:* In ultra-relativistic heavy-ion collisions, the transient magnetic field strength is expected to reach peak values of order of the squared pion mass. This makes it conceivable that strong electro-magnetic fields notably affect particle production processes in QCD. Most importantly, when one couples QCD fluid dynamics to QED, the QCD chiral anomaly is known to give rise to anomalous hydrodynamic phenomena that are fluid dynamic manifestations of the non-abelian quantum nature of QCD. Fluid dynamic codes that include coupling to such anomalous currents ("QCD magneto-hydrodynamics") are under development. Related phenomena include currents that flow along the direction of magnetic fields or intrinsic vorticity. Experimental signatures of such macroscopic manifestations of a quantum anomaly have been searched for in various charge-separation effects, but confounding hadronic effects complicate the interpretation of existing data. The fundamental interest in identifying such quantum properties of QCD fluidity provides strong motivation for seeking further constraints from complementary measurements.

**In summary**

In recent years, experimental advances in ultra-relativistic heavy-ion collisions and in the study of smaller collision systems have given experimental access to fundamental questions about the emergence of apparent collectivity in the soft physics sector of media controlled by the strong interaction, and in the microscopic mechanisms underlying it. Here, we have assessed the different facets of this problem with emphasis on the current experimental and theoretical state of the art and the fundamental questions raised by it. In particular, we have emphasized that additional experimental efforts to scrutinize the perfect-fluid paradigm, paralleled by further theoretical developments, have the potential of experimentally testing a more complete QCD transport theory that incorporates fluid dynamic behavior. Such a program of investigation should be able to delineate the limitations of the fluid dynamic picture in very small systems, at very early times due to strong initial gradients, and at late times due non-equilibrium kinetic processes close to the final freeze-out point. This would lead to a far more complete picture of dense partonic QCD, in which experimental constraints on fluid transport coefficients are complemented by experimentally testable information on the microscopic degrees of freedom of medium constituents and their elementary dynamics and thermodynamics. While there are still important uncertainties about the techniques and the accuracy with which this program can be completed, we believe that the present write-up documents is a step forward in identifying the central questions and relating them to future experiments.

**Bibliography**

Given the generality of the arguments in this document, any referencing of specific developments would be either incomplete or of incommensurate length. Therefore, the participants have agreed to present their contribution without referencing. Readers interested in learning more about the current state of the art in high-energy nuclear collisions are referred to [1] and references therein.

[1] Quark Matter 2015, https://indico.cern.ch/event/355454/, proceedings to be published in a special issue of Nuclear Physics A.